\documentclass[prl,aps,twocolumn,a4paper]{revtex4-1}
\usepackage{bm}
\usepackage{graphicx}
\usepackage{amsmath}
\usepackage{bm}
\usepackage{graphicx}
\usepackage{subfigure}
\usepackage{amsmath}

\begin{document}

\title{Phototaxis beyond turning: persistent accumulation and response acclimation of the microalga {\it Chlamydomonas reinhardtii}}

\author{Jorge Arrieta$^1$, Ana Barreira$^1$, Maurizio Chioccioli$^2$, Marco Polin$^3$, and Idan Tuval$^1$}
\affiliation{$^1$Mediterranean Institute for Advanced Studies (CSIC-UIB), Spain\\
$^2$Cavendish Laboratory, University of Cambridge, Cambridge CB3 0HE, United Kingdom\\
$^3$Physics Department, University of Warwick, Gibbet Hill Road, Coventry CV4 7AL, United Kingdom}


\begin{abstract}
Phototaxis is an important reaction to light  displayed by a wide range of motile microorganisms. Flagellated eukaryotic microalgae in particular, like the model organism {\it Chlamydomonas reinhardtii}, steer either towards or away from light by a rapid and precisely timed modulation of their flagellar activity. Cell steering, however, is only the beginning of a much longer process which ultimately allows cells to determine their light exposure history. This process is not well understood. Here we present a first quantitative study of the long timescale phototactic motility of {\it Chlamydomonas} at both single cell and population levels.
Our results reveal that the phototactic strategy adopted by these microorganisms leads to an efficient exposure to light, and that the phototactic response is modulated over typical timescales of tens of seconds.
The adaptation dynamics for phototaxis and chlorophyll fluorescence show a striking quantitative agreement, suggesting that photosynthesis controls quantitatively how cells navigate a light field.
\end{abstract}

\maketitle

The fitness of microorganisms depends critically on their ability to sense dynamic physico-chemical clues from the environment, elaborate the information and respond effectively. Environmental responses range from changes in gene expression \cite{Trippens2012} (typical timescale $\sim10\,$min); to the activation/deactivation of biochemical processes like chloroplast photoprotection \cite{Allorent2013} ($\sim1\,$min); to fast movement regulation ($\sim1\,$s), either active \cite{Stocker2008,Drescher2010} or passive \cite{Arrieta2015}. 
The best characterised motile response is currently chemotaxis of run-and-tumble bacteria like {\it E. coli} \cite{Berg1975}, a strategy based on the modulation of  tumbling frequency \cite{Celani2010}. Chemotaxis features (almost)perfect adaptation to persistent stimuli over intermediate timescales ($\sim10-100\,$s) \cite{Sourjik2002,Meir2010}  and can stimulate/inhibit gene expression through a variety of chemosensory pathways \cite{Wadhams2004}. This paradigmatic sensory system highlights the important crosstalk happening between responses acting across a wide spectrum of time intervals, and exemplifies the need for a consistent cross-timescale framework to understand motility regulation in microorganisms. In the case of phototaxis, a major response in eukaryotic microalgae \cite{Jekely2009}, this framework is lacking.

\begin{figure}[bt!]
   \centering
   \includegraphics[width=0.95\columnwidth]{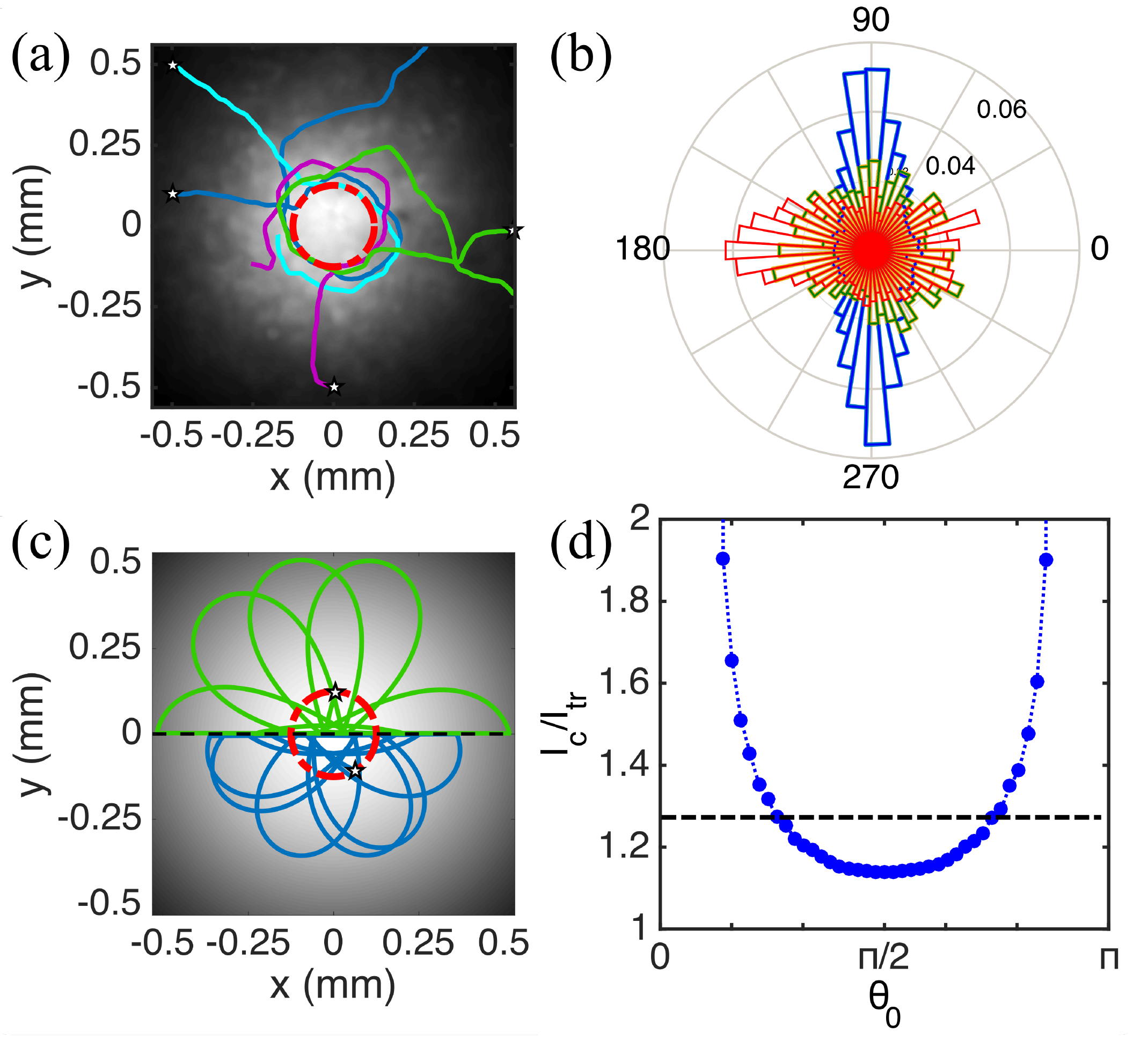} 
   \caption{Single cell phototaxis of {\it C. reinhardtii}. (a) Sample trajectories, starting at the star-marked points. Cells approach the centre of the light field, circulate around it at an average distance $\rho_c$ marked by the red dashed circle, and then leave the field of view. (b) Experimental histograms of cells' directions at $\rho = 78\,\mu$m (green), $156\,\mu$m (blue), and  $780\,\mu$m (red). The angle is oriented radially outwards. (c) Representative trajectories from local gradient model, with $\alpha=\alpha_{\textrm{max}}$, starting at  $\rho=\rho_c$ with initial orientations $\theta = 205^{\circ}$ (blue) and $162^{\circ}$ (green). For clarity, only half of each trajectory is displayed. Red dashed circle has radius $\rho_c$. The underlying light field is the best Gaussian fit to the experimental one. (d) Ratio between the average light intensity seen by a swimmer circulating at $\rho=\rho_c$ and moving along trochoidal trajectories starting at $(\rho_c,\theta_0)$ (blue circles; dotted line: guide to the eye). Black dashed line: average value of the relative increase in irradiance ($29\%$).}
   \label{fig1}
\end{figure}

Among micro-eukaryotes, phototaxis is best characterised in the model system {\it Chlamydomonas reinhardtii} (CR) \cite{Harris2009}, a green microalga which swims along a helical trajectory by the synchronous breaststroke beating of its flagellar pair \cite{Goldstein2009,Goldstein2011}. Cell spinning \cite{Martinez2012} induces a periodic modulation of the signal received by the eyespot, a rhodopsin-based light-sensitive organelle \cite{Kateriya2004} featuring a contrast-enhancing dielectric mirror \cite{Foster1980}. 
Eyespot stimulation is rapidly relayed {\it via} an action-potential-like signal to the flagella (ms)\cite{Sineshchekov1990}, and triggers a Ca$^{+2}$-dependent differential response of their beating \cite{Ruffer1991,Josef2006} causing cells to steer either towards or away from light \cite{Schaller1997,Yoshimura2001}. 
Implementation within a minimal model  \cite{Bennett2014} confirmed that phototactic steering is robust and can indeed lead to both positive and negative taxis, a property that has been used to achieve photo-hydrodynamic focussing of microalgae \cite{Garcia2013}. 
What happens beyond phototactic steering, however, is not well understood. Phototaxis of microalgae can lead to persistent modification of bioconvective patterns \cite{Williams2011a,Williams2011b}, and should therefore contribute to the interplay between fluid flow and motility leading to microscale patchiness in the seas  \cite{Torney2008,Stocker2012}. 
At the single cell level, phototaxis will modulate cell irradiance and can therefore be expected to impact both cell metabolism -through chloroplast stimulation- and light-sensitive gene expression \cite{Petroutsos2016}. Except for qualitative accounts of red-light control of phototactic sign \cite{Takahashi1993}, these links have not been explored.

Here we present the first long-timescale study of phototactic behaviour of CR, as representative of green microalgae. Studying the accumulation dynamics around a localised source, we show that cells use tight circulation around the maximum light intensity as a strategy to maximise their overall light exposure before spontaneously leaving the illuminated region. Periodic exposure experiments reveal that this is accompanied by a decrease in the overall response to light stimuli. The quantitative modulation of phototactic response tracks the dynamics of chlorophyll fluorescence, used here as a proxy for the photosynthetic activity of the cells.

\section*{Material and Methods}
{\it Chlamydomonas reinhardtii} wild type strain CC125 and bald mutant CC2905 were grown axenically at $24^{\circ}$C in Tris-Acetate-Phosphate medium \cite{TAP} under fluorescent light illumination (OSRAM Fluora, $100\,\mu$mol/m$^2$s PAR) following a 14h/10h light/dark diurnal cycle. Exponentially growing cells at $\sim2\times10^6\,$cells/ml were loaded in the  $7\,$mm diameter circular observation chamber cored out of a $1\,$mm thick agar pad sandwiched between coverslips. 
A CCD camera (Pike, AVT) hosted on a continuously focusable objective (InfiniVar CFM-2S, Infinity USA) recorded at $12.2\,$fps the phototactic motility of cells within the horizontal sample, visualised  through darkfield illumination at $635\,$nm (FLDR-i70A-R24, Falcon Lighting). Actinic light was provided by a $470\,$nm LED (Thorlabs M470L2) through a $200\,\mu$m-diameter multimode optical fibre (FT200EMT, Thorlabs). Approximation of the fibre output $I(\mathbf{x})$ by a Gaussian ($\sigma_I = 667 \,\mu$m, peak intensity $260\,\mu$mol/m$^2$s) is excellent and will be used throughout the paper. 
An inverted microscope (TE2000-U, Nikon) fitted with a $10\times$ Plan Apo objective (NA 0.45) and a EMCCD (Evolve, Photometrics) was used to record the chlorophyll fluorescence of CC2905, excited by the epiport-coupled blue LED.

\section*{Results and Discussion}
We begin by examining single-cell phototaxis after the light was kept on for $>10\,$min to ensure steady conditions (Fig.~\ref{fig1}a). Cells further from the centre than $200\,\mu$m move inwards along almost radial trajectories as a result of active steering. As they approach the centre, however, individual CRs turn sharply and start circulating around the maximum at an average distance of $\rho_c=139\pm 24\,\mu$m. This is confirmed by the azimuthally-averaged probability distribution function of swimming directions in Fig.~\ref{fig1}b. Given the average swimming speed $v_s=78\pm 11\,\mu$m/s, we obtain an angular velocity $\omega_c = 0.56\pm 0.125\,$rad/s which compares well with the average value previously reported for sharp turns ($\omega_m \simeq 0.8\,$rad/s) where cells achieve their largest angular speeds \cite{Polin2009a}. Orbiting cells do not show the preference for a particular chirality characteristic of hydrodynamic interactions with the sample surface \cite{Frymier1995,Lauga2006}. Instead, the orbits have a fundamentally phototactic origin. Recorded only episodically in flagellates \cite{Rhiel1988, Figueroa1998,Matsunaga2003}, orientation perpendicular to light stimulus (diaphototaxis) was reported as an anecdotal curiosity in CR \cite{Foster1980}. It appears here as a specific modulation of phototaxis allowing CR to dwell in localised light spots. 

The position $\mathbf{x}(t)$ of a cell swimming at constant speed $v_s$ along the direction $\mathbf{p}(t)$ will evolve according to 
\begin{equation}
\dot{\mathbf{x}}(t) = v_s\mathbf{p}(t)\quad;\quad \dot{\mathbf{p}}(t) = \bm{\omega}\wedge\mathbf{p}(t),
\label{eq:generic}
\end{equation}
where the angular speed $\bm{\omega}$ encodes the phototactic response through its (unknown) dependence on the light field. 
Absent detailed measurements, a common approach \cite{Williams2011a,Williams2011b,Furlan2012} has been to assume proportionality to the local gradient in light intensity, $\bm{\omega} = \alpha \mathbf{p}(t)\wedge\nabla I$, where the phototactic parameter $\alpha$, possibly dependent on $I$, represents the magnitude of the response.

For CR, the requirement $\omega\leq\omega_m$ implies $\alpha \leq \alpha_{\textrm{max}}=\omega_m/|\nabla I|_{\textrm{max}} $. This reasonable model predicts correctly the radial reorientation of cells far from the source, but the incoming trajectories are then expected to overshoot the centre and eventually describe trochoids like to those seen in Fig.~\ref{fig1}c. 
Similar trajectories are indeed seen both in phototactic colloids moving around a diverging laser beam \cite{Moyses2016}, and in sea-urchin sperm swimming around a local chemotactic cue \cite{Guerrero2010}. Phototactic CR's however, do not follow trochoids but fall instead onto the tightest closed loops they can achieve around the light source, at an average distance $\rho_c$ from the centre. This dynamics cannot be reproduced by changing $\alpha$ to include a transition between positive and negative phototaxis around $\rho_c$ ({\it SI Appendix}, Fig.~S1): it is a fundamentally different type of behaviour that cells follow during positive phototaxis.

\begin{figure*}[t]
   \begin{center}
   \includegraphics[width=0.9\textwidth]{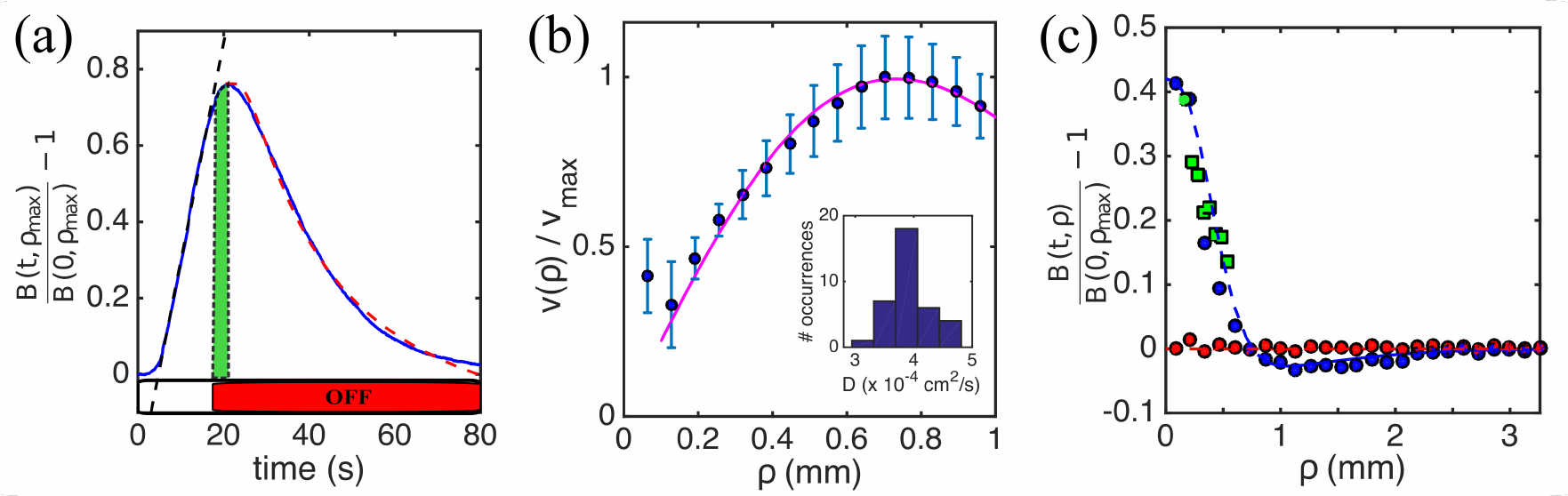} 
   \caption{Steady phototactic response of a population of {\it C. reinhardtii}. (a) Representative phototactic accumulation curve at $\rho=958\,\mu$m (blue solid line) as the phototactic light is turned on (at $t=0\,$s) and then off (at $t=15\,$s) as indicated by the coloured bars. Cells accumulate linearly (black dashed line: linear fit; slope $0.057\,\%$increase/s) and disperse diffusively (magenta dashed line: fit to diffusively spreading Gaussian). The green bar highlights the overshoot after light-off. (b) Average normalised phototactic velocity vs. distance from the fibre centre from $36$ different cycles. Errorbars: standard deviation of the measurement set. Magenta solid line: normalised light intensity gradient. The experimental light intensity is represented here by its best Gaussian fit. Inset: Effective diffusivities $D$ measured from $36$ different Gaussian fits to the dispersal curves.  (c) Radial concentration profiles from population experiments. Red circles: without light stimulus; blue circles: $35\,$s after light-on; green squares: concentration profile estimated using individual tracks from single-cell experiments; dashed blue line: one-parameter fit to the continuum model, giving $h^*=519\pm 27\,\mu$m.}
   \label{fig2}
   \end{center}
\end{figure*}

Fig.~\ref{fig1}d shows that the real dynamics exposes the microalgae to a $\sim 30\%$ larger path-averaged light intensity than the trochoidal case, and therefore appears to be better strategy to optimise light capture by a photosynthetic microswimmer. 
After $\tau_c=11.2\pm 2.5\,$s, however, cells stop orbiting and leave the field of view. Consistently observed across the 3290 tracks recorded, this behaviour reflects a clear adaptation of phototactic motility, turning here from positive to negative, and challenge the common assumption of an optimal light intensity which cells simply get attracted to. Flagellar response to light-step-up/step-down stimuli is indeed known to depend -qualitatively- on the choice of pre-stimulus adaptation  \cite{Ruffer1991}. The adaptive dynamics observed here, however, is a consequence of a history of light-exposure selected autonomously by single cells through their motility. 

We now turn to the phototactic behaviour of a population (Fig.~\ref{fig2}) to investigate the effect of adaptation over timescales longer than those accessible from the limited field of view of single-cell experiments. Cell concentrations will be kept below $5\times10^{6}\,$cells/ml to prevent effects on either the actinic light field perceived by the algae, or  darkfield illumination \cite{Schaller1997}. The image brightness $b(\mathbf{x},t)$ is then proportional to the 2D-projected concentration of algae $c(\mathbf{x},t)$, which integrates the 3D one across approximately the depth of field of the imaging apparatus. Agreement between brightness profiles after prolonged light exposure ($>35\,$s), and the distribution of cell positions from individual tracks (Fig.~\ref{fig2}c) confirms the proportionality, and suggests that cell-cell interactions are not important here. Cell accumulation can be characterised through the integrated image brightness $B(t; \rho) = 2\pi\int^{\rho} b(\rho,t)\rho\, \textrm{d}\rho$ where the maximum value $\rho_{max}=958\,\mu$m is set by the image size. 
Initially uniformly distributed, the algae begin to accumulate around the fibre as the light is turned on, causing $B(t;\rho)$ to increase linearly with time (Fig.~\ref{fig2}a, blue solid line). This is a signature of a constant inward flux of cells,  proportional to the product $\rho^2v_p(\rho)$ of the net phototactic drift $v_p(\rho)$ at distance $\rho$, and the geometric factor $\rho^2$ which takes into account cells moving inwards from deep within the sample. 
The full curve $v_p(\rho)$ can then be measured from the initial increase up to a multiplicative constant (Fig.~\ref{fig2}a, black dashed line). Figure~\ref{fig2}b shows that this is well described by $v_p(\rho)\propto |\nabla I|$ with the exception of the core region $\rho\lesssim 150\,\mu$m, where we already know that cell behaviour is different.

Switching the light off, the profile relaxes down to the original homogeneous value (Fig.~\ref{fig2}a, blue solid line). This dynamics is well characterised by a simple diffusive spreading (Fig.~\ref{fig2}b, magenta dashed line) with an effective diffusivity $D$ which can be recovered from a one-parameter fit (Fig.~\ref{fig2}b inset). The average value $\langle D\rangle = (3.9\pm0.4)\times10^{-4}\,$cm$^2$/s is in reasonable agreement with the average diffusivity $(4.7\pm0.5)\times10^{-4}\,$cm$^2$/s reported previously \cite{Polin2009a}.

The coarse-grained phototactic drift and the effective diffusivity can be used in a Keller-Segel-like continuum model of the phototactic behaviour of a population of CR, in the spirit of previous effective descriptions of phototaxis \cite{Furlan2012,Giometto2015,Martin2016}. 
In this model, valid  sufficiently far from the source, the local concentration of cells $c(\rho,t)$ moving in the fibre's axisymmetric light field $I(\rho)$ obeys the continuity equation 
\begin{equation}
\frac{\partial c}{\partial t} = \frac{\partial}{\partial\rho}\left(D\frac{\partial c}{\partial\rho} - c\frac{\rho}{h^*}\,v_p(\rho)\right),
\label{eq:concentration}
\end{equation}
where the extra factor of $\rho$, non-dimensionalised by the effective thickness $h^*$ has been included to take into account three dimensional effects on our 2D description, as discussed previously. The local phototactic velocity $v_p(\rho) = \beta\,v_s(\partial I(\rho)/\partial\rho)/(\partial I(\rho)/\partial\rho)_{\textrm{max}}$, which incorporates Weber's law \cite{Shoval2010}, is characterised by the phototactic sensitivity of the population, $\beta$, setting the maximum phototactic drift ($\beta \,v_s$). 
To compare Eq.~\eqref{eq:concentration} with  experiments, we fix the cell concentration at the sample boundary, $c(\mathbf{x},t)|_{\textrm{boundary}}=1$, and use the experimentally measured values for the mean swimming velocity and cell diffusivity, light field and phototactic sensitivity. The last parameter is derived from the distribution of single cells' swimming directions at $\rho=\sigma_I$ (Fig.~\ref{fig1}b), giving $\beta=0.14\pm0.013$. 
A one-parameter fit to the long-timescale profile in Fig.~\ref{fig2}c (blue circles) sets the value of $h^*$. 
The result (dashed blue line) shows that $h^* = 519\pm 27\,\mu$m provides an excellent description of the cell concentration, implying that cells within roughly half of the sample thickness take part in the phototactic accumulation.
The model predicts also the presence of a depletion ring at $\rho\simeq1.1\,$mm responsible for the slight overshoot of $B(t; \rho)$ experimentally observed right after light-off (Fig.~\ref{fig2}a, green bar). 
Single cell experiments suggest, then, that the measured low phototactic sensitivity results from the balance between inwards/outwards swimming and dwell time, all present in the natural phototactic behaviour of each individual CR cell and modulated by its irradiation history (Fig.~\ref{fig1}a).
\begin{figure}[t]
   \centering
   \includegraphics[width=0.95\columnwidth]{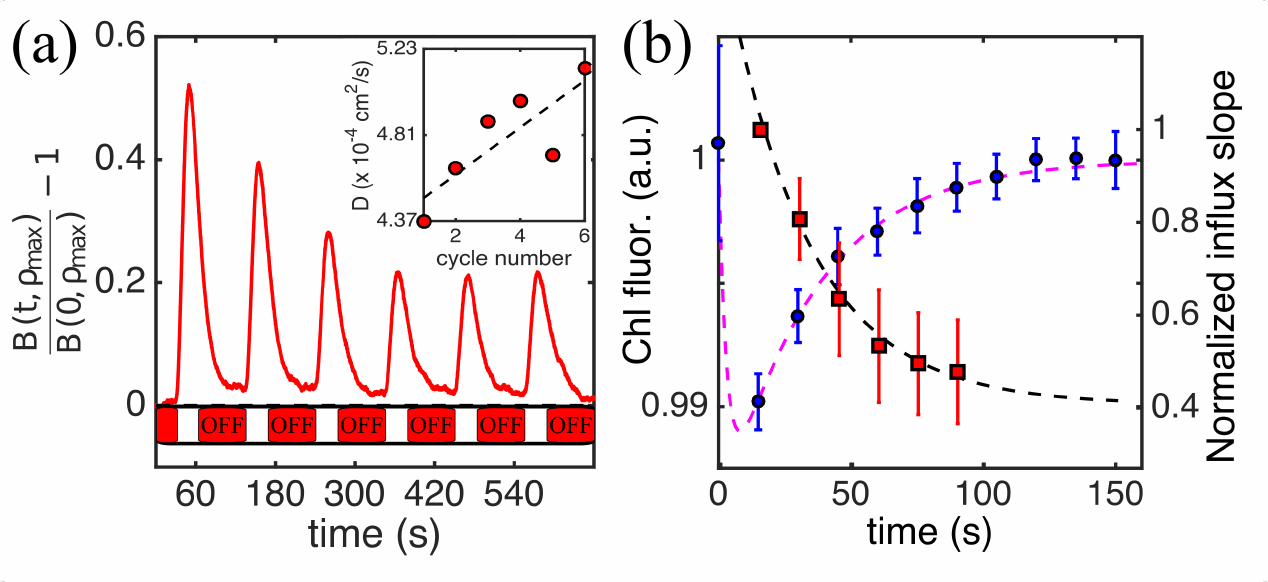} 
      \caption{Acclimation of the phototactic response. a) Representative accumulation and dispersal curves at $\rho=958\,\mu$m for six consecutive light on-off cycles. b) Red squares:  decay of the normalised phototactic sensitivity $\beta(t)/\beta(0)$ through the cycles. The time axis includes only periods of light-on. Error bars represent the standard deviation of the whole set of $60$ measurements. Black dashed line: exponential fit, giving an acclimation timescale of $\tau_{\beta} = 31.84\pm 1.94\,$s. Blue circles: evolution of the normalised chlorophyll fluorescence $\Phi_{\textrm{chl}}(t)/\Phi_{\textrm{chl}}(0)$ for CC2905 cells subjected to the same light on-off protocol. Error bars are the standard deviation of the whole set of $46$ repeats, each including $\sim1500$ cells on average. Magenta dashed line: fit to a two-timescale process. The initial fast response and the ensuing long acclimation are characterised respectively by the timescales $\tau^f_{\textrm{chl}}=1.47\pm 0.21\,$s and $\tau^s_{\textrm{chl}} = 33.49 \pm 5.2\,$s.
}
   \label{fig3}
\end{figure}

Equipped with an appropriate description of the steady state, we now investigate the adaptation process by characterising the phototactic accumulation of a population of dark-adapted cells to a series of identical light-on/light-off cycles ($15/90\,$s on/off; Movie S1 shows one cycle). Figure~\ref{fig3}a presents the accumulation dynamics for a representative experiment out of 60, showing a clear dependence on history of light exposure.
Accumulation and dispersal phases allow one to measure the time (and light) evolution of both $\beta$ and $D$, and therefore pinpoint the dynamical features responsible for the adaptation. 
Figure~\ref{fig3}a (inset) shows that over the whole experiment $D$ increases slightly by $\sim 15\%$,  suggesting a $\sim 7\%$ increase in $v_s$  (i.e. photokynesis) which, by itself, would lead to an equivalent increase in $\beta$.
Instead, this  parameter displays a well defined decrease through the cycles (Fig.~\ref{fig3}b, red squares), unequivocally assigning the adaptation to a change in the phototactic sensitivity alone. The evolution of the sensitivity parameter is well described by a single-time adaptation $\partial_t \beta(t) = \left( \beta^*-\beta(t) \right)/\tau_{\beta}$
where the adaptation timescale $\tau_{\beta}= 31.84\pm 1.94\,$s and $\beta^*/\beta(0) = 0.46\pm 0.19$  are derived from the fit in Fig.~\ref{fig3}b (black dashed line). In this analysis, we  assumed that $\beta$ evolves only during periods of illumination. 
Dark re-adaptation was not observed in the experiments; it must happen over significantly longer timescales and therefore was not considered here. 

Phototactic adaptation operates on timescales clearly separated from those characterising adaptation of either flagellar photoshock ($\sim1\,$s) \cite{Hegemann1989,Kateriya2004} or eyespot signalling ($\sim100\,$ms) \cite{Govorunova1997}. Being directly related to cell irradiance, itself relevant for photosynthesis, we therefore wondered whether the dynamics of $\beta$ would contain any  signature of light-adaptation by CR's photosynthetic apparatus. To investigate this, we exposed $\sim1500$ dark adapted non-swimming CR cells (CC2905) to the sequence of light stimulation used previously (see Fig.~\ref{fig3}a), and recorded the evolution of their average chlorophyll fluorescence $\Phi_{\textrm{chl}}$ ($502\,$nm$<\lambda<538\,$nm), which can be used as a simple proxy for the activity of the photosynthetic apparatus \cite{Maxwell2000}. 

A homogeneous light field of intensity $540\,\mu$E/m$^2$s was used (identical results were obtained for $975$ and $1320\,\mu$mol/m$^2$s). Figure~\ref{fig3}b shows the evolution of the mean $\Phi_{\textrm{chl}}(t)$ during each light-on period (blue circles). Light-off intervals did not induce appreciable dark-adaptation, in line with known differences between light- and dark-adaptation of the photosynthetic apparatus \cite{Baker2008}. Chlorophyll fluorescence evolution is well fitted by a simple two-timescale dynamics  (Fig.~\ref{fig3}b magenta dashed line) with an initial fast response (timescale $\tau^{f}_{\textrm{chl}} = 1.47\pm 0.21\,$s) followed by a slow adaptation with timescale $\tau^{s}_{\textrm{chl}} = 33.49\pm 5.2\,$s. The exceptional quantitative agreement between $\tau^{s}_{\textrm{chl}}$ and $\tau_{\beta}$ suggests a connection between the two processes, a possibility which would also explain the slow dark-adaptation of phototaxis. 

Phototaxis experiments under a simultaneous background illumination have shown that chloroplast stimulation can induce CRs to qualitatively switch their phototactic sign (positive to negative) \cite{Takahashi1993}. Our results suggest the intriguing possibility that phototaxis and photosynthesis are in fact connected quantitatively. Although further experiments are needed to firmly establish this layer of control, we propose here the hypothesis that this connection is indeed the major determinant of the phototactic motility of eukaryotic microalgae.

\section*{Conclusions}
The light-induced steering responses evolved by microorganisms like {\it Chlamydomonas} are complex, and have been studied extensively. Ultimately, however, flagellar activity must be integrated into a coherent navigation strategy combining physical stimuli and intracellular requirements: how this is achieved is currently not  understood. By shifting the focus to long timescales we start addressing this gap. Our experiments have already revealed a surprisingly rich dynamics, from the ability to increase light exposure by switching to diaphototaxis to the adaptive response of cells which reproduces the slow (re)adaptation of their chlorophyll fluorescence. Future experiments will be needed to systematically explore the role of light intensity and colour; to determine whether phototaxis shares any of the common properties of cellular sensory systems, like exact adaptation \cite{Shoval2010,Lazova2011}; and in particular how these properties are connected with photoprotective dynamics within  the chloroplast  \cite{Allorent2013}  and photosynthetic efficiency \cite{Kim2016}.

The authors acknowledge fruitful discussions with Miguel Gonzalez. The work has been partly supported by the Spanish Ministry of Economy and Competitiveness grant No. FIS2013-48444-C2-1-P and the Subprogram Ram\'on-y-Cajal (IT); the Royal Society Research Grant RG150421 and the University of the Balearic Islands Travel Grant 22/2016 (MP). MP gratefully acknowledges the hospitality of the Mediterranean Institute of Advanced Studies, where part of this work has been performed.

\bibliography{photoadaptation}

\begin{thebibliography}{49}%
\makeatletter
\providecommand \@ifxundefined [1]{%
 \@ifx{#1\undefined}
}%
\providecommand \@ifnum [1]{%
 \ifnum #1\expandafter \@firstoftwo
 \else \expandafter \@secondoftwo
 \fi
}%
\providecommand \@ifx [1]{%
 \ifx #1\expandafter \@firstoftwo
 \else \expandafter \@secondoftwo
 \fi
}%
\providecommand \natexlab [1]{#1}%
\providecommand \enquote  [1]{``#1''}%
\providecommand \bibnamefont  [1]{#1}%
\providecommand \bibfnamefont [1]{#1}%
\providecommand \citenamefont [1]{#1}%
\providecommand \href@noop [0]{\@secondoftwo}%
\providecommand \href [0]{\begingroup \@sanitize@url \@href}%
\providecommand \@href[1]{\@@startlink{#1}\@@href}%
\providecommand \@@href[1]{\endgroup#1\@@endlink}%
\providecommand \@sanitize@url [0]{\catcode `\\12\catcode `\$12\catcode
  `\&12\catcode `\#12\catcode `\^12\catcode `\_12\catcode `\%12\relax}%
\providecommand \@@startlink[1]{}%
\providecommand \@@endlink[0]{}%
\providecommand \url  [0]{\begingroup\@sanitize@url \@url }%
\providecommand \@url [1]{\endgroup\@href {#1}{\urlprefix }}%
\providecommand \urlprefix  [0]{URL }%
\providecommand \Eprint [0]{\href }%
\providecommand \doibase [0]{http://dx.doi.org/}%
\providecommand \selectlanguage [0]{\@gobble}%
\providecommand \bibinfo  [0]{\@secondoftwo}%
\providecommand \bibfield  [0]{\@secondoftwo}%
\providecommand \translation [1]{[#1]}%
\providecommand \BibitemOpen [0]{}%
\providecommand \bibitemStop [0]{}%
\providecommand \bibitemNoStop [0]{.\EOS\space}%
\providecommand \EOS [0]{\spacefactor3000\relax}%
\providecommand \BibitemShut  [1]{\csname bibitem#1\endcsname}%
\let\auto@bib@innerbib\@empty
\bibitem [{\citenamefont {Trippens}\ \emph {et~al.}(2012)\citenamefont
  {Trippens}, \citenamefont {Greiner}, \citenamefont {Schellwat}, \citenamefont
  {Neukam}, \citenamefont {Rottmann}, \citenamefont {Lu}, \citenamefont
  {Kateriya}, \citenamefont {Hegemann},\ and\ \citenamefont
  {Kreimer}}]{Trippens2012}%
  \BibitemOpen
  \bibfield  {author} {\bibinfo {author} {\bibfnamefont {J.}~\bibnamefont
  {Trippens}}, \bibinfo {author} {\bibfnamefont {A.}~\bibnamefont {Greiner}},
  \bibinfo {author} {\bibfnamefont {J.}~\bibnamefont {Schellwat}}, \bibinfo
  {author} {\bibfnamefont {M.}~\bibnamefont {Neukam}}, \bibinfo {author}
  {\bibfnamefont {T.}~\bibnamefont {Rottmann}}, \bibinfo {author}
  {\bibfnamefont {Y.}~\bibnamefont {Lu}}, \bibinfo {author} {\bibfnamefont
  {S.}~\bibnamefont {Kateriya}}, \bibinfo {author} {\bibfnamefont
  {P.}~\bibnamefont {Hegemann}}, \ and\ \bibinfo {author} {\bibfnamefont
  {G.}~\bibnamefont {Kreimer}},\ }\href {\doibase 10.1105/tpc.112.103523}
  {\bibfield  {journal} {\bibinfo  {journal} {The Plant cell}\ }\textbf
  {\bibinfo {volume} {24}},\ \bibinfo {pages} {4687} (\bibinfo {year}
  {2012})}\BibitemShut {NoStop}%
\bibitem [{\citenamefont {Allorent}\ \emph {et~al.}(2013)\citenamefont
  {Allorent}, \citenamefont {Tokutsu}, \citenamefont {Roach}, \citenamefont
  {Peers}, \citenamefont {Cardol}, \citenamefont {Girard-Bascou}, \citenamefont
  {Seigneurin-Berny}, \citenamefont {Petroutsos}, \citenamefont {Kuntz},
  \citenamefont {Breyton}, \citenamefont {Franck}, \citenamefont {Wollman},
  \citenamefont {Niyogi}, \citenamefont {Krieger-Liszkay}, \citenamefont
  {Minagawa},\ and\ \citenamefont {Finazzi}}]{Allorent2013}%
  \BibitemOpen
  \bibfield  {author} {\bibinfo {author} {\bibfnamefont {G.}~\bibnamefont
  {Allorent}}, \bibinfo {author} {\bibfnamefont {R.}~\bibnamefont {Tokutsu}},
  \bibinfo {author} {\bibfnamefont {T.}~\bibnamefont {Roach}}, \bibinfo
  {author} {\bibfnamefont {G.}~\bibnamefont {Peers}}, \bibinfo {author}
  {\bibfnamefont {P.}~\bibnamefont {Cardol}}, \bibinfo {author} {\bibfnamefont
  {J.}~\bibnamefont {Girard-Bascou}}, \bibinfo {author} {\bibfnamefont
  {D.}~\bibnamefont {Seigneurin-Berny}}, \bibinfo {author} {\bibfnamefont
  {D.}~\bibnamefont {Petroutsos}}, \bibinfo {author} {\bibfnamefont
  {M.}~\bibnamefont {Kuntz}}, \bibinfo {author} {\bibfnamefont
  {C.}~\bibnamefont {Breyton}}, \bibinfo {author} {\bibfnamefont
  {F.}~\bibnamefont {Franck}}, \bibinfo {author} {\bibfnamefont {F.-a.}\
  \bibnamefont {Wollman}}, \bibinfo {author} {\bibfnamefont {K.~K.}\
  \bibnamefont {Niyogi}}, \bibinfo {author} {\bibfnamefont {A.}~\bibnamefont
  {Krieger-Liszkay}}, \bibinfo {author} {\bibfnamefont {J.}~\bibnamefont
  {Minagawa}}, \ and\ \bibinfo {author} {\bibfnamefont {G.}~\bibnamefont
  {Finazzi}},\ }\href {\doibase 10.1105/tpc.112.108274} {\bibfield  {journal}
  {\bibinfo  {journal} {The Plant Cell}\ }\textbf {\bibinfo {volume} {25}},\
  \bibinfo {pages} {545} (\bibinfo {year} {2013})}\BibitemShut {NoStop}%
\bibitem [{\citenamefont {Stocker}\ \emph {et~al.}(2008)\citenamefont
  {Stocker}, \citenamefont {Seymour}, \citenamefont {Samadani}, \citenamefont
  {Hunt},\ and\ \citenamefont {Polz}}]{Stocker2008}%
  \BibitemOpen
  \bibfield  {author} {\bibinfo {author} {\bibfnamefont {R.}~\bibnamefont
  {Stocker}}, \bibinfo {author} {\bibfnamefont {J.~R.}\ \bibnamefont
  {Seymour}}, \bibinfo {author} {\bibfnamefont {A.}~\bibnamefont {Samadani}},
  \bibinfo {author} {\bibfnamefont {D.~E.}\ \bibnamefont {Hunt}}, \ and\
  \bibinfo {author} {\bibfnamefont {M.~F.}\ \bibnamefont {Polz}},\ }\href
  {\doibase 10.1073/pnas.0709765105} {\bibfield  {journal} {\bibinfo  {journal}
  {Proceedings of the National Academy of Sciences}\ }\textbf {\bibinfo
  {volume} {105}},\ \bibinfo {pages} {4209} (\bibinfo {year}
  {2008})}\BibitemShut {NoStop}%
\bibitem [{\citenamefont {Drescher}\ \emph {et~al.}(2010)\citenamefont
  {Drescher}, \citenamefont {Goldstein},\ and\ \citenamefont
  {Tuval}}]{Drescher2010}%
  \BibitemOpen
  \bibfield  {author} {\bibinfo {author} {\bibfnamefont {K.}~\bibnamefont
  {Drescher}}, \bibinfo {author} {\bibfnamefont {R.~E.}\ \bibnamefont
  {Goldstein}}, \ and\ \bibinfo {author} {\bibfnamefont {I.}~\bibnamefont
  {Tuval}},\ }\href {\doibase 10.1073/pnas.1000901107} {\bibfield  {journal}
  {\bibinfo  {journal} {Proceedings of the National Academy of Sciences of the
  United States of America}\ }\textbf {\bibinfo {volume} {107}},\ \bibinfo
  {pages} {11171} (\bibinfo {year} {2010})}\BibitemShut {NoStop}%
\bibitem [{\citenamefont {Arrieta}\ \emph {et~al.}(2015)\citenamefont
  {Arrieta}, \citenamefont {Barreira},\ and\ \citenamefont
  {Tuval}}]{Arrieta2015}%
  \BibitemOpen
  \bibfield  {author} {\bibinfo {author} {\bibfnamefont {J.}~\bibnamefont
  {Arrieta}}, \bibinfo {author} {\bibfnamefont {A.}~\bibnamefont {Barreira}}, \
  and\ \bibinfo {author} {\bibfnamefont {I.}~\bibnamefont {Tuval}},\ }\href
  {\doibase 10.1103/PhysRevLett.114.128102} {\bibfield  {journal} {\bibinfo
  {journal} {Physical Review Letters}\ }\textbf {\bibinfo {volume} {114}},\
  \bibinfo {pages} {128102} (\bibinfo {year} {2015})}\BibitemShut {NoStop}%
\bibitem [{\citenamefont {Berg}(1975)}]{Berg1975}%
  \BibitemOpen
  \bibfield  {author} {\bibinfo {author} {\bibfnamefont {H.~C.}\ \bibnamefont
  {Berg}},\ }\href {\doibase 10.1146/annurev.bb.04.060175.001003} {\bibfield
  {journal} {\bibinfo  {journal} {Annual review of biophysics and
  bioengineering}\ }\textbf {\bibinfo {volume} {4}},\ \bibinfo {pages} {119}
  (\bibinfo {year} {1975})}\BibitemShut {NoStop}%
\bibitem [{\citenamefont {Celani}\ and\ \citenamefont
  {Vergassola}(2010)}]{Celani2010}%
  \BibitemOpen
  \bibfield  {author} {\bibinfo {author} {\bibfnamefont {A.}~\bibnamefont
  {Celani}}\ and\ \bibinfo {author} {\bibfnamefont {M.}~\bibnamefont
  {Vergassola}},\ }\href {\doibase 10.1073/pnas.0909673107} {\bibfield
  {journal} {\bibinfo  {journal} {Proceedings of the National Academy of
  Sciences}\ }\textbf {\bibinfo {volume} {107}},\ \bibinfo {pages} {1391}
  (\bibinfo {year} {2010})}\BibitemShut {NoStop}%
\bibitem [{\citenamefont {Sourjik}\ and\ \citenamefont
  {Berg}(2002)}]{Sourjik2002}%
  \BibitemOpen
  \bibfield  {author} {\bibinfo {author} {\bibfnamefont {V.}~\bibnamefont
  {Sourjik}}\ and\ \bibinfo {author} {\bibfnamefont {H.~C.}\ \bibnamefont
  {Berg}},\ }\href {\doibase 10.1073/pnas.011589998} {\bibfield  {journal}
  {\bibinfo  {journal} {Proceedings of the National Academy of Sciences}\
  }\textbf {\bibinfo {volume} {99}},\ \bibinfo {pages} {123} (\bibinfo {year}
  {2002})}\BibitemShut {NoStop}%
\bibitem [{\citenamefont {Meir}\ \emph {et~al.}(2010)\citenamefont {Meir},
  \citenamefont {Jakovljevic}, \citenamefont {Oleksiuk}, \citenamefont
  {Sourjik},\ and\ \citenamefont {Wingreen}}]{Meir2010}%
  \BibitemOpen
  \bibfield  {author} {\bibinfo {author} {\bibfnamefont {Y.}~\bibnamefont
  {Meir}}, \bibinfo {author} {\bibfnamefont {V.}~\bibnamefont {Jakovljevic}},
  \bibinfo {author} {\bibfnamefont {O.}~\bibnamefont {Oleksiuk}}, \bibinfo
  {author} {\bibfnamefont {V.}~\bibnamefont {Sourjik}}, \ and\ \bibinfo
  {author} {\bibfnamefont {N.~S.}\ \bibnamefont {Wingreen}},\ }\href {\doibase
  10.1016/j.bpj.2010.08.051} {\bibfield  {journal} {\bibinfo  {journal}
  {Biophysical Journal}\ }\textbf {\bibinfo {volume} {99}},\ \bibinfo {pages}
  {2766} (\bibinfo {year} {2010})}\BibitemShut {NoStop}%
\bibitem [{\citenamefont {Wadhams}\ and\ \citenamefont
  {Armitage}(2004)}]{Wadhams2004}%
  \BibitemOpen
  \bibfield  {author} {\bibinfo {author} {\bibfnamefont {G.~H.}\ \bibnamefont
  {Wadhams}}\ and\ \bibinfo {author} {\bibfnamefont {J.~P.}\ \bibnamefont
  {Armitage}},\ }\href {\doibase 10.1038/nrm1524} {\bibfield  {journal}
  {\bibinfo  {journal} {Nature Reviews Molecular Cell Biology}\ }\textbf
  {\bibinfo {volume} {5}},\ \bibinfo {pages} {1024} (\bibinfo {year}
  {2004})}\BibitemShut {NoStop}%
\bibitem [{\citenamefont {Jekely}(2009)}]{Jekely2009}%
  \BibitemOpen
  \bibfield  {author} {\bibinfo {author} {\bibfnamefont {G.}~\bibnamefont
  {Jekely}},\ }\href {\doibase 10.1098/rstb.2009.0072} {\bibfield  {journal}
  {\bibinfo  {journal} {Philosophical Transactions of the Royal Society B:
  Biological sciences}\ }\textbf {\bibinfo {volume} {364}},\ \bibinfo {pages}
  {2795} (\bibinfo {year} {2009})}\BibitemShut {NoStop}%
\bibitem [{\citenamefont {Harris}(2009)}]{Harris2009}%
  \BibitemOpen
  \bibfield  {author} {\bibinfo {author} {\bibfnamefont {E.~H.}\ \bibnamefont
  {Harris}},\ }\href@noop {} {\emph {\bibinfo {title} {{The Chalmydomonas
  Sourcebook}}}}\ (\bibinfo {year} {2009})\BibitemShut {NoStop}%
\bibitem [{\citenamefont {Goldstein}\ \emph {et~al.}(2009)\citenamefont
  {Goldstein}, \citenamefont {Polin},\ and\ \citenamefont
  {Tuval}}]{Goldstein2009}%
  \BibitemOpen
  \bibfield  {author} {\bibinfo {author} {\bibfnamefont {R.~E.}\ \bibnamefont
  {Goldstein}}, \bibinfo {author} {\bibfnamefont {M.}~\bibnamefont {Polin}}, \
  and\ \bibinfo {author} {\bibfnamefont {I.}~\bibnamefont {Tuval}},\ }\href
  {\doibase 10.1103/PhysRevLett.103.168103} {\bibfield  {journal} {\bibinfo
  {journal} {Physical Review Letters}\ }\textbf {\bibinfo {volume} {103}},\
  \bibinfo {pages} {168103} (\bibinfo {year} {2009})}\BibitemShut {NoStop}%
\bibitem [{\citenamefont {Goldstein}\ \emph {et~al.}(2011)\citenamefont
  {Goldstein}, \citenamefont {Polin},\ and\ \citenamefont
  {Tuval}}]{Goldstein2011}%
  \BibitemOpen
  \bibfield  {author} {\bibinfo {author} {\bibfnamefont {R.~E.}\ \bibnamefont
  {Goldstein}}, \bibinfo {author} {\bibfnamefont {M.}~\bibnamefont {Polin}}, \
  and\ \bibinfo {author} {\bibfnamefont {I.}~\bibnamefont {Tuval}},\ }\href
  {\doibase 10.1103/PhysRevLett.107.148103} {\bibfield  {journal} {\bibinfo
  {journal} {Physical Review Letters}\ }\textbf {\bibinfo {volume} {107}},\
  \bibinfo {pages} {148103} (\bibinfo {year} {2011})}\BibitemShut {NoStop}%
\bibitem [{\citenamefont {Martinez}\ \emph {et~al.}(2012)\citenamefont
  {Martinez}, \citenamefont {Besseling}, \citenamefont {Croze}, \citenamefont
  {Tailleur}, \citenamefont {Reufer}, \citenamefont {Schwarz-Linek},
  \citenamefont {Wilson}, \citenamefont {Bees},\ and\ \citenamefont
  {Poon}}]{Martinez2012}%
  \BibitemOpen
  \bibfield  {author} {\bibinfo {author} {\bibfnamefont {V.~a.}\ \bibnamefont
  {Martinez}}, \bibinfo {author} {\bibfnamefont {R.}~\bibnamefont {Besseling}},
  \bibinfo {author} {\bibfnamefont {O.~a.}\ \bibnamefont {Croze}}, \bibinfo
  {author} {\bibfnamefont {J.}~\bibnamefont {Tailleur}}, \bibinfo {author}
  {\bibfnamefont {M.}~\bibnamefont {Reufer}}, \bibinfo {author} {\bibfnamefont
  {J.}~\bibnamefont {Schwarz-Linek}}, \bibinfo {author} {\bibfnamefont {L.~G.}\
  \bibnamefont {Wilson}}, \bibinfo {author} {\bibfnamefont {M.~A.}\
  \bibnamefont {Bees}}, \ and\ \bibinfo {author} {\bibfnamefont {W.~C.~K.}\
  \bibnamefont {Poon}},\ }\href {\doibase 10.1016/j.bpj.2012.08.045} {\bibfield
   {journal} {\bibinfo  {journal} {Biophysical journal}\ }\textbf {\bibinfo
  {volume} {103}},\ \bibinfo {pages} {1637} (\bibinfo {year}
  {2012})}\BibitemShut {NoStop}%
\bibitem [{\citenamefont {Kateriya}(2004)}]{Kateriya2004}%
  \BibitemOpen
  \bibfield  {author} {\bibinfo {author} {\bibfnamefont {S.}~\bibnamefont
  {Kateriya}},\ }\href {\doibase 10.1152/nips.01517.2004} {\bibfield  {journal}
  {\bibinfo  {journal} {News in Physiological Sciences}\ }\textbf {\bibinfo
  {volume} {19}},\ \bibinfo {pages} {133} (\bibinfo {year} {2004})}\BibitemShut
  {NoStop}%
\bibitem [{\citenamefont {Foster}\ and\ \citenamefont
  {Smyth}(1980)}]{Foster1980}%
  \BibitemOpen
  \bibfield  {author} {\bibinfo {author} {\bibfnamefont {K.~W.}\ \bibnamefont
  {Foster}}\ and\ \bibinfo {author} {\bibfnamefont {R.~D.}\ \bibnamefont
  {Smyth}},\ }\href@noop {} {\bibfield  {journal} {\bibinfo  {journal}
  {Microbiological Review}\ }\textbf {\bibinfo {volume} {44}},\ \bibinfo
  {pages} {572} (\bibinfo {year} {1980})}\BibitemShut {NoStop}%
\bibitem [{\citenamefont {Sineshchekov}\ \emph {et~al.}(1990)\citenamefont
  {Sineshchekov}, \citenamefont {Litvin},\ and\ \citenamefont
  {Keszthelyi}}]{Sineshchekov1990}%
  \BibitemOpen
  \bibfield  {author} {\bibinfo {author} {\bibfnamefont {O.~a.}\ \bibnamefont
  {Sineshchekov}}, \bibinfo {author} {\bibfnamefont {F.~F.}\ \bibnamefont
  {Litvin}}, \ and\ \bibinfo {author} {\bibfnamefont {L.}~\bibnamefont
  {Keszthelyi}},\ }\href {\doibase 10.1016/S0006-3495(90)82504-2} {\bibfield
  {journal} {\bibinfo  {journal} {Biophysical Journal}\ }\textbf {\bibinfo
  {volume} {57}},\ \bibinfo {pages} {33} (\bibinfo {year} {1990})}\BibitemShut
  {NoStop}%
\bibitem [{\citenamefont {Ruffer}\ and\ \citenamefont
  {Nultsch}(1991)}]{Ruffer1991}%
  \BibitemOpen
  \bibfield  {author} {\bibinfo {author} {\bibfnamefont {U.}~\bibnamefont
  {Ruffer}}\ and\ \bibinfo {author} {\bibfnamefont {W.}~\bibnamefont
  {Nultsch}},\ }\href@noop {} {\bibfield  {journal} {\bibinfo  {journal} {Cell
  Motility and the Cytoskeleton}\ }\textbf {\bibinfo {volume} {18}},\ \bibinfo
  {pages} {269} (\bibinfo {year} {1991})}\BibitemShut {NoStop}%
\bibitem [{\citenamefont {Josef}\ \emph {et~al.}(2006)\citenamefont {Josef},
  \citenamefont {Saranak},\ and\ \citenamefont {Foster}}]{Josef2006}%
  \BibitemOpen
  \bibfield  {author} {\bibinfo {author} {\bibfnamefont {K.}~\bibnamefont
  {Josef}}, \bibinfo {author} {\bibfnamefont {J.}~\bibnamefont {Saranak}}, \
  and\ \bibinfo {author} {\bibfnamefont {K.~W.}\ \bibnamefont {Foster}},\
  }\href {\doibase 10.1002/cm.20158} {\bibfield  {journal} {\bibinfo  {journal}
  {Cell motility and the cytoskeleton}\ }\textbf {\bibinfo {volume} {63}},\
  \bibinfo {pages} {758} (\bibinfo {year} {2006})}\BibitemShut {NoStop}%
\bibitem [{\citenamefont {Schaller}\ \emph {et~al.}(1997)\citenamefont
  {Schaller}, \citenamefont {David},\ and\ \citenamefont {Uhl}}]{Schaller1997}%
  \BibitemOpen
  \bibfield  {author} {\bibinfo {author} {\bibfnamefont {K.}~\bibnamefont
  {Schaller}}, \bibinfo {author} {\bibfnamefont {R.}~\bibnamefont {David}}, \
  and\ \bibinfo {author} {\bibfnamefont {R.}~\bibnamefont {Uhl}},\ }\href
  {\doibase 10.1016/S0006-3495(97)78188-8} {\bibfield  {journal} {\bibinfo
  {journal} {Biophysical Journal}\ }\textbf {\bibinfo {volume} {73}},\ \bibinfo
  {pages} {1562} (\bibinfo {year} {1997})}\BibitemShut {NoStop}%
\bibitem [{\citenamefont {Yoshimura}\ and\ \citenamefont
  {Kamiya}(2001)}]{Yoshimura2001}%
  \BibitemOpen
  \bibfield  {author} {\bibinfo {author} {\bibfnamefont {K.}~\bibnamefont
  {Yoshimura}}\ and\ \bibinfo {author} {\bibfnamefont {R.}~\bibnamefont
  {Kamiya}},\ }\href@noop {} {\bibfield  {journal} {\bibinfo  {journal} {Plant
  and Cell Physiology}\ }\textbf {\bibinfo {volume} {42}},\ \bibinfo {pages}
  {665} (\bibinfo {year} {2001})}\BibitemShut {NoStop}%
\bibitem [{\citenamefont {Bennett}\ and\ \citenamefont
  {Golestanian}(2014)}]{Bennett2014}%
  \BibitemOpen
  \bibfield  {author} {\bibinfo {author} {\bibfnamefont {R.~R.}\ \bibnamefont
  {Bennett}}\ and\ \bibinfo {author} {\bibfnamefont {R.}~\bibnamefont
  {Golestanian}},\ }\href {\doibase 10.1098/rsif.2014.1164} {\bibfield
  {journal} {\bibinfo  {journal} {Journal of the Royal Society Interface}\
  }\textbf {\bibinfo {volume} {12}},\ \bibinfo {pages} {20141164} (\bibinfo
  {year} {2014})},\ \Eprint {http://arxiv.org/abs/1410.6270} {arXiv:1410.6270}
  \BibitemShut {NoStop}%
\bibitem [{\citenamefont {Garcia}\ \emph {et~al.}(2013)\citenamefont {Garcia},
  \citenamefont {Rafa{\"{i}}},\ and\ \citenamefont {Peyla}}]{Garcia2013}%
  \BibitemOpen
  \bibfield  {author} {\bibinfo {author} {\bibfnamefont {X.}~\bibnamefont
  {Garcia}}, \bibinfo {author} {\bibfnamefont {S.}~\bibnamefont {Rafa{\"{i}}}},
  \ and\ \bibinfo {author} {\bibfnamefont {P.}~\bibnamefont {Peyla}},\ }\href
  {\doibase 10.1103/PhysRevLett.110.138106} {\bibfield  {journal} {\bibinfo
  {journal} {Physical Review Letters}\ }\textbf {\bibinfo {volume} {110}},\
  \bibinfo {pages} {138106} (\bibinfo {year} {2013})}\BibitemShut {NoStop}%
\bibitem [{\citenamefont {Williams}\ and\ \citenamefont
  {Bees}(2011{\natexlab{a}})}]{Williams2011a}%
  \BibitemOpen
  \bibfield  {author} {\bibinfo {author} {\bibfnamefont {C.~R.}\ \bibnamefont
  {Williams}}\ and\ \bibinfo {author} {\bibfnamefont {M.~A.}\ \bibnamefont
  {Bees}},\ }\href {\doibase 10.1242/jeb.051094} {\bibfield  {journal}
  {\bibinfo  {journal} {The Journal of experimental biology}\ }\textbf
  {\bibinfo {volume} {214}},\ \bibinfo {pages} {2398} (\bibinfo {year}
  {2011}{\natexlab{a}})}\BibitemShut {NoStop}%
\bibitem [{\citenamefont {Williams}\ and\ \citenamefont
  {Bees}(2011{\natexlab{b}})}]{Williams2011b}%
  \BibitemOpen
  \bibfield  {author} {\bibinfo {author} {\bibfnamefont {C.~R.}\ \bibnamefont
  {Williams}}\ and\ \bibinfo {author} {\bibfnamefont {M.~A.}\ \bibnamefont
  {Bees}},\ }\href {\doibase 10.1017/jfm.2011.100} {\bibfield  {journal}
  {\bibinfo  {journal} {Journal of Fluid Mechanics}\ }\textbf {\bibinfo
  {volume} {678}},\ \bibinfo {pages} {41} (\bibinfo {year}
  {2011}{\natexlab{b}})}\BibitemShut {NoStop}%
\bibitem [{\citenamefont {Torney}\ and\ \citenamefont
  {Neufeld}(2008)}]{Torney2008}%
  \BibitemOpen
  \bibfield  {author} {\bibinfo {author} {\bibfnamefont {C.}~\bibnamefont
  {Torney}}\ and\ \bibinfo {author} {\bibfnamefont {Z.}~\bibnamefont
  {Neufeld}},\ }\href {\doibase 10.1103/PhysRevLett.101.078105} {\bibfield
  {journal} {\bibinfo  {journal} {Physical Review Letters}\ }\textbf {\bibinfo
  {volume} {101}},\ \bibinfo {pages} {1} (\bibinfo {year} {2008})}\BibitemShut
  {NoStop}%
\bibitem [{\citenamefont {Stocker}(2012)}]{Stocker2012}%
  \BibitemOpen
  \bibfield  {author} {\bibinfo {author} {\bibfnamefont {R.}~\bibnamefont
  {Stocker}},\ }\href {\doibase 10.1126/science.1208929} {\bibfield  {journal}
  {\bibinfo  {journal} {Science}\ }\textbf {\bibinfo {volume} {338}},\ \bibinfo
  {pages} {628} (\bibinfo {year} {2012})}\BibitemShut {NoStop}%
\bibitem [{\citenamefont {Petroutsos}\ \emph {et~al.}(2016)\citenamefont
  {Petroutsos}, \citenamefont {Tokutsu}, \citenamefont {Maruyama},
  \citenamefont {Flori}, \citenamefont {Greiner}, \citenamefont {Magneschi},
  \citenamefont {Cusant}, \citenamefont {Kottke}, \citenamefont {Mittag},
  \citenamefont {Hegemann}, \citenamefont {Finazzi},\ and\ \citenamefont
  {Minagawa}}]{Petroutsos2016}%
  \BibitemOpen
  \bibfield  {author} {\bibinfo {author} {\bibfnamefont {D.}~\bibnamefont
  {Petroutsos}}, \bibinfo {author} {\bibfnamefont {R.}~\bibnamefont {Tokutsu}},
  \bibinfo {author} {\bibfnamefont {S.}~\bibnamefont {Maruyama}}, \bibinfo
  {author} {\bibfnamefont {S.}~\bibnamefont {Flori}}, \bibinfo {author}
  {\bibfnamefont {A.}~\bibnamefont {Greiner}}, \bibinfo {author} {\bibfnamefont
  {L.}~\bibnamefont {Magneschi}}, \bibinfo {author} {\bibfnamefont
  {L.}~\bibnamefont {Cusant}}, \bibinfo {author} {\bibfnamefont
  {T.}~\bibnamefont {Kottke}}, \bibinfo {author} {\bibfnamefont
  {M.}~\bibnamefont {Mittag}}, \bibinfo {author} {\bibfnamefont
  {P.}~\bibnamefont {Hegemann}}, \bibinfo {author} {\bibfnamefont
  {G.}~\bibnamefont {Finazzi}}, \ and\ \bibinfo {author} {\bibfnamefont
  {J.}~\bibnamefont {Minagawa}},\ }\href {\doibase 10.1038/nature19358}
  {\bibfield  {journal} {\bibinfo  {journal} {Nature}\ }\textbf {\bibinfo
  {volume} {537}},\ \bibinfo {pages} {563} (\bibinfo {year}
  {2016})}\BibitemShut {NoStop}%
\bibitem [{\citenamefont {Takahashi}\ and\ \citenamefont
  {Watanabe}(1993)}]{Takahashi1993}%
  \BibitemOpen
  \bibfield  {author} {\bibinfo {author} {\bibfnamefont {T.}~\bibnamefont
  {Takahashi}}\ and\ \bibinfo {author} {\bibfnamefont {M.}~\bibnamefont
  {Watanabe}},\ }\href {\doibase 10.1016/0014-5793(93)80867-T} {\bibfield
  {journal} {\bibinfo  {journal} {FEBS Letters}\ }\textbf {\bibinfo {volume}
  {336}},\ \bibinfo {pages} {516} (\bibinfo {year} {1993})}\BibitemShut
  {NoStop}%
\bibitem [{\citenamefont {Rochaix}\ \emph {et~al.}(1988)\citenamefont
  {Rochaix}, \citenamefont {Mayfield}, \citenamefont {Goldschmidt-Clermont},\
  and\ \citenamefont {Erickson}}]{TAP}%
  \BibitemOpen
  \bibfield  {author} {\bibinfo {author} {\bibfnamefont {J.~D.}\ \bibnamefont
  {Rochaix}}, \bibinfo {author} {\bibfnamefont {S.}~\bibnamefont {Mayfield}},
  \bibinfo {author} {\bibfnamefont {M.}~\bibnamefont {Goldschmidt-Clermont}}, \
  and\ \bibinfo {author} {\bibfnamefont {J.~M.}\ \bibnamefont {Erickson}},\
  }\href@noop {} {\emph {\bibinfo {title} {{Plant Molecular Biology: A
  Practical Approach}}}},\ edited by\ \bibinfo {editor} {\bibfnamefont {C.~H.}\
  \bibnamefont {Schaw}}\ (\bibinfo  {publisher} {IRL Press, Oxford, England},\
  \bibinfo {year} {1988})\ \bibinfo {note} {pp. 253-275}\BibitemShut {NoStop}%
\bibitem [{\citenamefont {Polin}\ \emph {et~al.}(2009)\citenamefont {Polin},
  \citenamefont {Tuval}, \citenamefont {Drescher}, \citenamefont {Gollub},\
  and\ \citenamefont {Goldstein}}]{Polin2009a}%
  \BibitemOpen
  \bibfield  {author} {\bibinfo {author} {\bibfnamefont {M.}~\bibnamefont
  {Polin}}, \bibinfo {author} {\bibfnamefont {I.}~\bibnamefont {Tuval}},
  \bibinfo {author} {\bibfnamefont {K.}~\bibnamefont {Drescher}}, \bibinfo
  {author} {\bibfnamefont {J.~P.}\ \bibnamefont {Gollub}}, \ and\ \bibinfo
  {author} {\bibfnamefont {R.~E.}\ \bibnamefont {Goldstein}},\ }\href {\doibase
  10.1126/science.1172667} {\bibfield  {journal} {\bibinfo  {journal}
  {Science}\ }\textbf {\bibinfo {volume} {325}},\ \bibinfo {pages} {487}
  (\bibinfo {year} {2009})}\BibitemShut {NoStop}%
\bibitem [{\citenamefont {Frymier}\ \emph {et~al.}(1995)\citenamefont
  {Frymier}, \citenamefont {Ford}, \citenamefont {Berg},\ and\ \citenamefont
  {Cummings}}]{Frymier1995}%
  \BibitemOpen
  \bibfield  {author} {\bibinfo {author} {\bibfnamefont {P.~D.}\ \bibnamefont
  {Frymier}}, \bibinfo {author} {\bibfnamefont {R.~M.}\ \bibnamefont {Ford}},
  \bibinfo {author} {\bibfnamefont {H.~C.}\ \bibnamefont {Berg}}, \ and\
  \bibinfo {author} {\bibfnamefont {P.~T.}\ \bibnamefont {Cummings}},\ }\href
  {\doibase 10.1073/pnas.92.13.6195} {\bibfield  {journal} {\bibinfo  {journal}
  {Proceedings of the National Academy of Sciences of the United States of
  America}\ }\textbf {\bibinfo {volume} {92}},\ \bibinfo {pages} {6195}
  (\bibinfo {year} {1995})}\BibitemShut {NoStop}%
\bibitem [{\citenamefont {Lauga}\ \emph {et~al.}(2006)\citenamefont {Lauga},
  \citenamefont {DiLuzio}, \citenamefont {Whitesides},\ and\ \citenamefont
  {Stone}}]{Lauga2006}%
  \BibitemOpen
  \bibfield  {author} {\bibinfo {author} {\bibfnamefont {E.}~\bibnamefont
  {Lauga}}, \bibinfo {author} {\bibfnamefont {W.~R.}\ \bibnamefont {DiLuzio}},
  \bibinfo {author} {\bibfnamefont {G.~M.}\ \bibnamefont {Whitesides}}, \ and\
  \bibinfo {author} {\bibfnamefont {H.~A.}\ \bibnamefont {Stone}},\ }\href
  {\doibase 10.1529/biophysj.105.069401} {\bibfield  {journal} {\bibinfo
  {journal} {Biophysical Journal}\ }\textbf {\bibinfo {volume} {90}},\ \bibinfo
  {pages} {400} (\bibinfo {year} {2006})},\ \Eprint
  {http://arxiv.org/abs/0506675} {arXiv:0506675 [cond-mat]} \BibitemShut
  {NoStop}%
\bibitem [{\citenamefont {Rhiel}\ \emph {et~al.}(1988)\citenamefont {Rhiel},
  \citenamefont {Hader},\ and\ \citenamefont {Wehrmeyer}}]{Rhiel1988}%
  \BibitemOpen
  \bibfield  {author} {\bibinfo {author} {\bibfnamefont {E.}~\bibnamefont
  {Rhiel}}, \bibinfo {author} {\bibfnamefont {D.-p.}\ \bibnamefont {Hader}}, \
  and\ \bibinfo {author} {\bibfnamefont {W.}~\bibnamefont {Wehrmeyer}},\
  }\href@noop {} {\bibfield  {journal} {\bibinfo  {journal} {Plant and Cell
  Physiology}\ }\textbf {\bibinfo {volume} {29}},\ \bibinfo {pages} {755}
  (\bibinfo {year} {1988})}\BibitemShut {NoStop}%
\bibitem [{\citenamefont {Figueroa}\ \emph {et~al.}(1998)\citenamefont
  {Figueroa}, \citenamefont {Niell}, \citenamefont {Figueiras},\ and\
  \citenamefont {Villarino}}]{Figueroa1998}%
  \BibitemOpen
  \bibfield  {author} {\bibinfo {author} {\bibfnamefont {F.~L.}\ \bibnamefont
  {Figueroa}}, \bibinfo {author} {\bibfnamefont {F.~X.}\ \bibnamefont {Niell}},
  \bibinfo {author} {\bibfnamefont {F.~G.}\ \bibnamefont {Figueiras}}, \ and\
  \bibinfo {author} {\bibfnamefont {M.~L.}\ \bibnamefont {Villarino}},\
  }\href@noop {} {\bibfield  {journal} {\bibinfo  {journal} {Marine Biology}\
  }\textbf {\bibinfo {volume} {130}},\ \bibinfo {pages} {491} (\bibinfo {year}
  {1998})}\BibitemShut {NoStop}%
\bibitem [{\citenamefont {Matsunaga}\ \emph {et~al.}(2003)\citenamefont
  {Matsunaga}, \citenamefont {Watanabe}, \citenamefont {Sakaushi},
  \citenamefont {Miyamura},\ and\ \citenamefont {Hori}}]{Matsunaga2003}%
  \BibitemOpen
  \bibfield  {author} {\bibinfo {author} {\bibfnamefont {S.}~\bibnamefont
  {Matsunaga}}, \bibinfo {author} {\bibfnamefont {S.}~\bibnamefont {Watanabe}},
  \bibinfo {author} {\bibfnamefont {S.}~\bibnamefont {Sakaushi}}, \bibinfo
  {author} {\bibfnamefont {S.}~\bibnamefont {Miyamura}}, \ and\ \bibinfo
  {author} {\bibfnamefont {T.}~\bibnamefont {Hori}},\ }\href {\doibase
  10.1562/0031-8655(2003)077<0324:SEDTSD>2.0.CO;2} {\bibfield  {journal}
  {\bibinfo  {journal} {Photochemistry and Photobiology}\ }\textbf {\bibinfo
  {volume} {77}},\ \bibinfo {pages} {324} (\bibinfo {year} {2003})}\BibitemShut
  {NoStop}%
\bibitem [{\citenamefont {Furlan}\ \emph {et~al.}(2012)\citenamefont {Furlan},
  \citenamefont {Comparini}, \citenamefont {Ciszak}, \citenamefont {Beccai},
  \citenamefont {Mancuso},\ and\ \citenamefont {Mazzolai}}]{Furlan2012}%
  \BibitemOpen
  \bibfield  {author} {\bibinfo {author} {\bibfnamefont {S.}~\bibnamefont
  {Furlan}}, \bibinfo {author} {\bibfnamefont {D.}~\bibnamefont {Comparini}},
  \bibinfo {author} {\bibfnamefont {M.}~\bibnamefont {Ciszak}}, \bibinfo
  {author} {\bibfnamefont {L.}~\bibnamefont {Beccai}}, \bibinfo {author}
  {\bibfnamefont {S.}~\bibnamefont {Mancuso}}, \ and\ \bibinfo {author}
  {\bibfnamefont {B.}~\bibnamefont {Mazzolai}},\ }\href {\doibase
  10.1371/journal.pone.0038895} {\bibfield  {journal} {\bibinfo  {journal}
  {PloS one}\ }\textbf {\bibinfo {volume} {7}},\ \bibinfo {pages} {e38895}
  (\bibinfo {year} {2012})}\BibitemShut {NoStop}%
\bibitem [{\citenamefont {Moyses}\ \emph {et~al.}(2016)\citenamefont {Moyses},
  \citenamefont {Palacci}, \citenamefont {Sacanna},\ and\ \citenamefont
  {Grier}}]{Moyses2016}%
  \BibitemOpen
  \bibfield  {author} {\bibinfo {author} {\bibfnamefont {H.}~\bibnamefont
  {Moyses}}, \bibinfo {author} {\bibfnamefont {J.}~\bibnamefont {Palacci}},
  \bibinfo {author} {\bibfnamefont {S.}~\bibnamefont {Sacanna}}, \ and\
  \bibinfo {author} {\bibfnamefont {D.~G.}\ \bibnamefont {Grier}},\ }\href
  {\doibase 10.1039/C6SM01163B} {\bibfield  {journal} {\bibinfo  {journal}
  {Soft Matter}\ }\textbf {\bibinfo {volume} {12}},\ \bibinfo {pages} {6357}
  (\bibinfo {year} {2016})}\BibitemShut {NoStop}%
\bibitem [{\citenamefont {Guerrero}\ \emph {et~al.}(2010)\citenamefont
  {Guerrero}, \citenamefont {Nishigaki}, \citenamefont {Carneiro},
  \citenamefont {{Yoshiro Tatsu}}, \citenamefont {Wood},\ and\ \citenamefont
  {Darszon}}]{Guerrero2010}%
  \BibitemOpen
  \bibfield  {author} {\bibinfo {author} {\bibfnamefont {A.}~\bibnamefont
  {Guerrero}}, \bibinfo {author} {\bibfnamefont {T.}~\bibnamefont {Nishigaki}},
  \bibinfo {author} {\bibfnamefont {J.}~\bibnamefont {Carneiro}}, \bibinfo
  {author} {\bibnamefont {{Yoshiro Tatsu}}}, \bibinfo {author} {\bibfnamefont
  {C.~D.}\ \bibnamefont {Wood}}, \ and\ \bibinfo {author} {\bibfnamefont
  {A.}~\bibnamefont {Darszon}},\ }\href {\doibase 10.1016/j.ydbio.2010.04.013}
  {\bibfield  {journal} {\bibinfo  {journal} {Developmental Biology}\ }\textbf
  {\bibinfo {volume} {344}},\ \bibinfo {pages} {52} (\bibinfo {year}
  {2010})}\BibitemShut {NoStop}%
\bibitem [{\citenamefont {Giometto}\ \emph {et~al.}(2015)\citenamefont
  {Giometto}, \citenamefont {Altermatt}, \citenamefont {Maritan}, \citenamefont
  {Stocker},\ and\ \citenamefont {Rinaldo}}]{Giometto2015}%
  \BibitemOpen
  \bibfield  {author} {\bibinfo {author} {\bibfnamefont {A.}~\bibnamefont
  {Giometto}}, \bibinfo {author} {\bibfnamefont {F.}~\bibnamefont {Altermatt}},
  \bibinfo {author} {\bibfnamefont {A.}~\bibnamefont {Maritan}}, \bibinfo
  {author} {\bibfnamefont {R.}~\bibnamefont {Stocker}}, \ and\ \bibinfo
  {author} {\bibfnamefont {A.}~\bibnamefont {Rinaldo}},\ }\href {\doibase
  10.1073/pnas.1422922112} {\bibfield  {journal} {\bibinfo  {journal}
  {Proceedings of the National Academy of Sciences}\ }\textbf {\bibinfo
  {volume} {112}},\ \bibinfo {pages} {7045} (\bibinfo {year}
  {2015})}\BibitemShut {NoStop}%
\bibitem [{\citenamefont {Martin}\ \emph {et~al.}(2016)\citenamefont {Martin},
  \citenamefont {Barzyk}, \citenamefont {Bertin}, \citenamefont {Peyla},\ and\
  \citenamefont {Rafai}}]{Martin2016}%
  \BibitemOpen
  \bibfield  {author} {\bibinfo {author} {\bibfnamefont {M.}~\bibnamefont
  {Martin}}, \bibinfo {author} {\bibfnamefont {A.}~\bibnamefont {Barzyk}},
  \bibinfo {author} {\bibfnamefont {E.}~\bibnamefont {Bertin}}, \bibinfo
  {author} {\bibfnamefont {P.}~\bibnamefont {Peyla}}, \ and\ \bibinfo {author}
  {\bibfnamefont {S.}~\bibnamefont {Rafai}},\ }\href {\doibase
  10.1103/PhysRevE.93.051101} {\bibfield  {journal} {\bibinfo  {journal}
  {Physical Review E}\ }\textbf {\bibinfo {volume} {93}},\ \bibinfo {pages}
  {051101} (\bibinfo {year} {2016})},\ \Eprint
  {http://arxiv.org/abs/1603.00761} {arXiv:1603.00761} \BibitemShut {NoStop}%
\bibitem [{\citenamefont {Shoval}\ \emph {et~al.}(2010)\citenamefont {Shoval},
  \citenamefont {Goentoro}, \citenamefont {Hart}, \citenamefont {Mayo},
  \citenamefont {Sontag},\ and\ \citenamefont {Alon}}]{Shoval2010}%
  \BibitemOpen
  \bibfield  {author} {\bibinfo {author} {\bibfnamefont {O.}~\bibnamefont
  {Shoval}}, \bibinfo {author} {\bibfnamefont {L.}~\bibnamefont {Goentoro}},
  \bibinfo {author} {\bibfnamefont {Y.}~\bibnamefont {Hart}}, \bibinfo {author}
  {\bibfnamefont {A.}~\bibnamefont {Mayo}}, \bibinfo {author} {\bibfnamefont
  {E.}~\bibnamefont {Sontag}}, \ and\ \bibinfo {author} {\bibfnamefont
  {U.}~\bibnamefont {Alon}},\ }\href {\doibase 10.1073/pnas.1002352107}
  {\bibfield  {journal} {\bibinfo  {journal} {Proceedings of the National
  Academy of Sciences of the United States of America}\ }\textbf {\bibinfo
  {volume} {107}},\ \bibinfo {pages} {15995} (\bibinfo {year}
  {2010})}\BibitemShut {NoStop}%
\bibitem [{\citenamefont {Hegemann}\ and\ \citenamefont
  {Bruck}(1989)}]{Hegemann1989}%
  \BibitemOpen
  \bibfield  {author} {\bibinfo {author} {\bibfnamefont {P.}~\bibnamefont
  {Hegemann}}\ and\ \bibinfo {author} {\bibfnamefont {B.}~\bibnamefont
  {Bruck}},\ }\href@noop {} {\bibfield  {journal} {\bibinfo  {journal} {Cell
  Motility and the Cytoskeleton}\ }\textbf {\bibinfo {volume} {14}},\ \bibinfo
  {pages} {501} (\bibinfo {year} {1989})}\BibitemShut {NoStop}%
\bibitem [{\citenamefont {Govorunova}\ \emph {et~al.}(1997)\citenamefont
  {Govorunova}, \citenamefont {Sineshchekov},\ and\ \citenamefont
  {Hegemann}}]{Govorunova1997}%
  \BibitemOpen
  \bibfield  {author} {\bibinfo {author} {\bibfnamefont {E.~G.}\ \bibnamefont
  {Govorunova}}, \bibinfo {author} {\bibfnamefont {O.~A.}\ \bibnamefont
  {Sineshchekov}}, \ and\ \bibinfo {author} {\bibfnamefont {P.}~\bibnamefont
  {Hegemann}},\ }\href@noop {} {\bibfield  {journal} {\bibinfo  {journal}
  {Plant Physiology}\ }\textbf {\bibinfo {volume} {115}},\ \bibinfo {pages}
  {633} (\bibinfo {year} {1997})}\BibitemShut {NoStop}%
\bibitem [{\citenamefont {Maxwell}\ and\ \citenamefont
  {Johnson}(2000)}]{Maxwell2000}%
  \BibitemOpen
  \bibfield  {author} {\bibinfo {author} {\bibfnamefont {K.}~\bibnamefont
  {Maxwell}}\ and\ \bibinfo {author} {\bibfnamefont {G.~N.}\ \bibnamefont
  {Johnson}},\ }\href {http://www.ncbi.nlm.nih.gov/pubmed/10938857} {\bibfield
  {journal} {\bibinfo  {journal} {Journal of experimental botany}\ }\textbf
  {\bibinfo {volume} {51}},\ \bibinfo {pages} {659} (\bibinfo {year}
  {2000})}\BibitemShut {NoStop}%
\bibitem [{\citenamefont {Baker}(2008)}]{Baker2008}%
  \BibitemOpen
  \bibfield  {author} {\bibinfo {author} {\bibfnamefont {N.~R.}\ \bibnamefont
  {Baker}},\ }\href {\doibase 10.1146/annurev.arplant.59.032607.092759}
  {\bibfield  {journal} {\bibinfo  {journal} {Annual review of plant biology}\
  }\textbf {\bibinfo {volume} {59}},\ \bibinfo {pages} {89} (\bibinfo {year}
  {2008})}\BibitemShut {NoStop}%
\bibitem [{\citenamefont {Lazova}\ \emph {et~al.}(2011)\citenamefont {Lazova},
  \citenamefont {Ahmed}, \citenamefont {Bellomo}, \citenamefont {Stocker},\
  and\ \citenamefont {Shimizu}}]{Lazova2011}%
  \BibitemOpen
  \bibfield  {author} {\bibinfo {author} {\bibfnamefont {M.~D.}\ \bibnamefont
  {Lazova}}, \bibinfo {author} {\bibfnamefont {T.}~\bibnamefont {Ahmed}},
  \bibinfo {author} {\bibfnamefont {D.}~\bibnamefont {Bellomo}}, \bibinfo
  {author} {\bibfnamefont {R.}~\bibnamefont {Stocker}}, \ and\ \bibinfo
  {author} {\bibfnamefont {T.~S.}\ \bibnamefont {Shimizu}},\ }\href {\doibase
  10.1073/pnas.1108608108} {\bibfield  {journal} {\bibinfo  {journal}
  {Proceedings of the National Academy of Sciences}\ }\textbf {\bibinfo
  {volume} {108}},\ \bibinfo {pages} {13870} (\bibinfo {year}
  {2011})}\BibitemShut {NoStop}%
\bibitem [{\citenamefont {Kim}\ \emph {et~al.}(2016)\citenamefont {Kim},
  \citenamefont {Kwak}, \citenamefont {Sung}, \citenamefont {Choi},
  \citenamefont {Hong}, \citenamefont {Lim}, \citenamefont {Lee}, \citenamefont
  {Lee},\ and\ \citenamefont {Sim}}]{Kim2016}%
  \BibitemOpen
  \bibfield  {author} {\bibinfo {author} {\bibfnamefont {J.~Y.~H.}\
  \bibnamefont {Kim}}, \bibinfo {author} {\bibfnamefont {H.~S.}\ \bibnamefont
  {Kwak}}, \bibinfo {author} {\bibfnamefont {Y.~J.}\ \bibnamefont {Sung}},
  \bibinfo {author} {\bibfnamefont {H.~I.}\ \bibnamefont {Choi}}, \bibinfo
  {author} {\bibfnamefont {M.~E.}\ \bibnamefont {Hong}}, \bibinfo {author}
  {\bibfnamefont {H.~S.}\ \bibnamefont {Lim}}, \bibinfo {author} {\bibfnamefont
  {J.-H.}\ \bibnamefont {Lee}}, \bibinfo {author} {\bibfnamefont {S.~Y.}\
  \bibnamefont {Lee}}, \ and\ \bibinfo {author} {\bibfnamefont {S.~J.}\
  \bibnamefont {Sim}},\ }\href {\doibase 10.1038/srep21155} {\bibfield
  {journal} {\bibinfo  {journal} {Scientific Reports}\ }\textbf {\bibinfo
  {volume} {6}},\ \bibinfo {pages} {21155} (\bibinfo {year}
  {2016})}\BibitemShut {NoStop}%
\end{thebibliography}%

\setcounter{figure}{0}
\renewcommand{\thefigure}{S\arabic{figure}}
\section{Supplementary Informations}

\subsection*{Supplementary movie}The movie Mov1s.avi shows the dynamics of cell accumulation and dispersion in our experiments for a single light on - light off cycle.

\subsection*{Models for single cell phototaxis}
The position $\mathbf{x}(t)$ of a cell swimming at constant speed $v_s$ along the direction $\mathbf{p}(t)$ will evolve according to 
\begin{equation}
\dot{\mathbf{x}}(t) = v_s\mathbf{p}(t)\quad;\quad \dot{\mathbf{p}}(t) = \bm{\omega}\wedge\mathbf{p}(t),
\label{eq:generic}
\end{equation}
where the angular speed $\bm{\omega}$ encodes the phototactic response through its (unknown) dependence on the light field. 
Absent detailed measurements, a common approach has been to assume proportionality to the local gradient in light intensity, $\bm{\omega} = \alpha \mathbf{p}(t)\wedge\nabla I$, where the phototactic parameter $\alpha$, possibly dependent on $I$, represents the magnitude of the response.

\begin{figure*}[b]
   \begin{center}
   \includegraphics[width=0.7\textwidth]{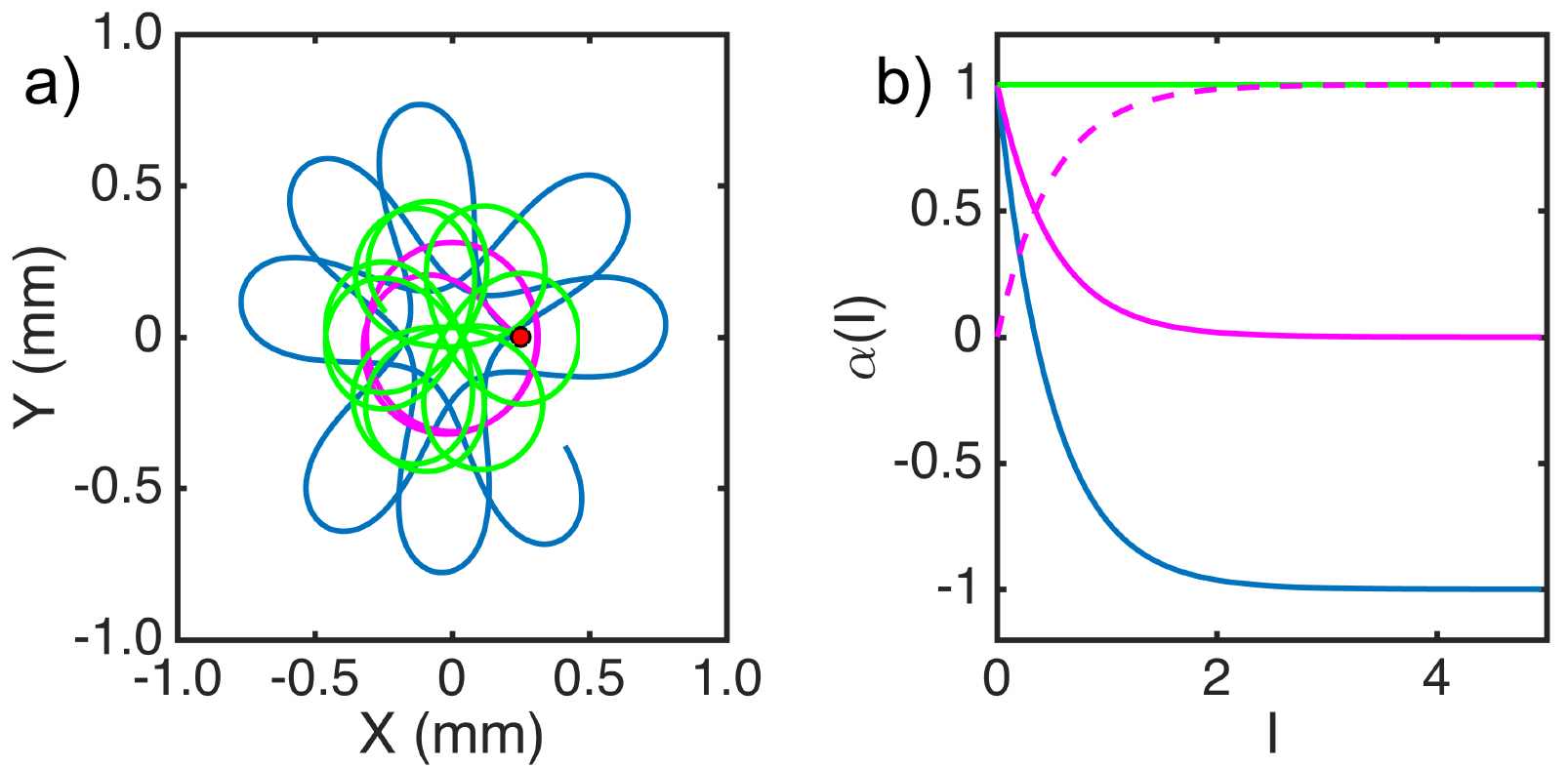}
   \caption{$\alpha(I)$ (a) and representative trajectories (b) for the three different individual based models explored. All the trajectories were initialized starting at  $\rho=\rho_c$ with initial orientations $\theta = 205^{\circ}$. The local gradient model used throughout the manuscript (green) as compared with a model that includes a transition from positive to negative phototaxis at a critical light intensity $I_c$ (blue) and with a phenomenological model incorporating diaphototaxis (magenta). The critical light intensity was set to $I_c=0.5$ in all cases.}
   \label{fig1s}
   \end{center}
\end{figure*}

The model used throughout the manuscript uses the simplest description that incorporates $\alpha(I)=\,$constant. Here we address two additional functional forms for $\alpha(I)$ chosen to model (a) the transition from positive to negative phototaxis and (b) the transition from positive to dia-phototaxis. For the former, we chose a continuous function of I that changes sign at a prescribed critical intensity; more specifically, we set $\alpha(I)=1-2 \exp(I/I_c)$ as seen in Fig. \ref{fig1s}a (blue curve). An example of a trajectory corresponding to this choice of $\alpha(I)$ is depicted with the same colour in panel (b). For the model to incorporate a transition to diaphototaxis it is necessary to include in the equation for $\bm{\omega}$ a term proportional to $\mathbf{p}(t)\cdot\nabla I$. In our case cells are confined to move on the $xy$ plane and therefore $\bm{\omega}=\omega\hat{e}_z$ where $\hat{e}_z$ is the unit vector in the direction perpendicular to the plane of motion. We have explored as a particular realisation of the phenomenological diaphototactic model the following:
\begin{equation}
\omega = \alpha(I) (\mathbf{p}(t)\wedge\nabla I)\cdot\hat{e}_z + (1-\alpha(I)) \mathbf{p}(t)\cdot\nabla I,
\label{eq:diaphototaxis}
\end{equation}
with $\alpha(I)=\exp(I/I_c)$. The model represents a continuous decay of positive phototaxis to zero for $I>I_c$ (shown as a magenta line in Fig. \ref{fig1s}a) and a concurrent increase of the diaphototactic contribution (dashed magenta line in the same figure). The combined effect of the two contributions leads to the cell circling around the position of $I=I_c$ as shown for the trajectory plotted with the same colour in panel (b).

\end{document}